\documentclass[10pt, twocolumn]{IEEEtran}
\IEEEoverridecommandlockouts

\usepackage{graphicx}
\usepackage{float}

\usepackage{cite}
\usepackage{amsmath,amssymb,amsfonts}
\usepackage{algorithmic}
\usepackage{diagbox}
\usepackage{algorithm}
\usepackage{multirow}
\usepackage{caption}
\usepackage{graphicx}
\usepackage{subfigure}
\usepackage{textcomp}
\usepackage{xcolor}
\usepackage{booktabs}
\def\BibTeX{{\rm B\kern-.05em{\sc i\kern-.025em b}\kern-.08em
		T\kern-.1667em\lower.7ex\hbox{E}\kern-.125emX}}
\begin{document}
\newtheorem{definition}{\it Definition}
\newtheorem{theorem}{\bf Theorem}
\newtheorem{lemma}{\it Lemma}
\newtheorem{corollary}{\it Corollary}
\newtheorem{remark}{\it Remark}
\newtheorem{example}{\it Example}
\newtheorem{case}{\bf Case Study}
\newtheorem{assumption}{\it Assumption}
\newtheorem{property}{\it Property}

\newtheorem{proposition}{\it Proposition}

\newcommand{\hP}[1]{{\boldsymbol h}_{{#1}{\bullet}}}
\newcommand{\hS}[1]{{\boldsymbol h}_{{\bullet}{#1}}}

\newcommand{\ba}{\boldsymbol{a}}
\newcommand{\baq}{\overline{q}}
\newcommand{\bA}{\boldsymbol{A}}
\newcommand{\bb}{\boldsymbol{b}}
\newcommand{\bB}{\boldsymbol{B}}
\newcommand{\bc}{\boldsymbol{c}}
\newcommand{\bp}{\boldsymbol{p}}
\newcommand{\bcO}{\boldsymbol{\cal O}}
\newcommand{\be}{\boldsymbol{e}}
\newcommand{\bh}{\boldsymbol{h}}
\newcommand{\bH}{\boldsymbol{H}}
\newcommand{\bl}{\boldsymbol{l}}
\newcommand{\bm}{\boldsymbol{m}}
\newcommand{\bn}{\boldsymbol{n}}
\newcommand{\bo}{\boldsymbol{o}}
\newcommand{\bO}{\boldsymbol{O}}
\newcommand{\bq}{\boldsymbol{q}}
\newcommand{\br}{\boldsymbol{r}}
\newcommand{\bR}{\boldsymbol{R}}
\newcommand{\bs}{\boldsymbol{s}}
\newcommand{\bS}{\boldsymbol{S}}
\newcommand{\bT}{\boldsymbol{T}}
\newcommand{\bw}{\boldsymbol{w}}

\newcommand{\balpha}{\boldsymbol{\alpha}}
\newcommand{\bbeta}{\boldsymbol{\beta}}
\newcommand{\bomega}{\boldsymbol{\omega}}
\newcommand{\bOmega}{\boldsymbol{\Omega}}
\newcommand{\bphi}{\boldsymbol{\phi}}
\newcommand{\bvarpi}{\boldsymbol{\varpi}}
\newcommand{\bpi}{\boldsymbol{\pi}}
\newcommand{\bxi}{\boldsymbol{\xi}}
\newcommand{\bx}{\boldsymbol{x}}
\newcommand{\by}{\boldsymbol{y}}

\newcommand{\cA}{{\cal A}}
\newcommand{\bcA}{\boldsymbol{\cal A}}
\newcommand{\cB}{{\cal B}}
\newcommand{\cE}{{\cal E}}
\newcommand{\cG}{{\cal G}}
\newcommand{\cH}{{\cal H}}
\newcommand{\bcH}{\boldsymbol {\cal H}}
\newcommand{\cK}{{\cal K}}
\newcommand{\cO}{{\cal O}}
\newcommand{\cR}{{\cal R}}
\newcommand{\cS}{{\cal S}}
\newcommand{\dcS}{\ddot{{\cal S}}}
\newcommand{\ds}{\ddot{{s}}}
\newcommand{\cT}{{\cal T}}
\newcommand{\cU}{{\cal U}}
\newcommand{\wt}[1]{\widetilde{#1}}

\newcommand{\mA}{\mathbb{A}}
\newcommand{\mE}{\mathbb{E}}
\newcommand{\mG}{\mathbb{G}}
\newcommand{\mR}{\mathbb{R}}
\newcommand{\mS}{\mathbb{S}}
\newcommand{\mU}{\mathbb{U}}
\newcommand{\mV}{\mathbb{V}}
\newcommand{\mW}{\mathbb{W}}

\newcommand{\uq}{\underline{q}}
\newcommand{\ubq}{\underline{\boldsymbol q}}

\newcommand{\red}[1]{\textcolor[rgb]{1,0,0}{#1}}
\newcommand{\gre}[1]{\textcolor[rgb]{0,1,0}{#1}}
\newcommand{\blu}[1]{\textcolor[rgb]{0,0,1}{#1}}

	\title{Reasoning on the Air: An Implicit Semantic Communication Architecture}
	\author{\IEEEauthorblockA{Yong~Xiao\IEEEauthorrefmark{1}\IEEEauthorrefmark{4}, Yingyu Li\IEEEauthorrefmark{3},  Guangming~Shi\IEEEauthorrefmark{2}\IEEEauthorrefmark{4}, and H. Vincent Poor\IEEEauthorrefmark{5} \\
\IEEEauthorblockA{\IEEEauthorrefmark{1}School of Elect. Inform. \& Commun., Huazhong Univ. of Science \& Technology, Wuhan, China}\\
\IEEEauthorblockA{\IEEEauthorrefmark{2}School of Mech. Eng. and Elect. Inform., China University of Geosciences, Wuhan, China}\\
\IEEEauthorblockA{\IEEEauthorrefmark{3}School of Artificial Intelligence, Xidian University, Xi'an, China}\\
\IEEEauthorblockA{\IEEEauthorrefmark{4}Pengcheng National Laboratory (Guangzhou base), Guangzhou, China}\\
\IEEEauthorblockA{\IEEEauthorrefmark{5}School of Engineering \& Applied Science, Princeton University, Princeton, NJ}\\
}
\thanks{This paper is accepted at IEEE ICC Workshop, Seoul, South Korea, May 2022.}
}

\maketitle

\begin{abstract}
Semantic communication is a novel communication paradigm which draws inspiration from human communication focusing on the delivery of the meaning of a message to the intended users. It has attracted significant interest recently due to its potential to improve efficiency and reliability of communication, enhance users' quality-of-experience (QoE), and achieve smoother cross-protocol/domain communication. Most existing works in semantic communication focus on identifying and transmitting explicit semantic meaning, e.g., labels of objects, that can be directly identified from the source signal. This paper investigates implicit semantic communication in which the hidden information, e.g., implicit causality and reasoning mechanisms of users, that cannot be directly observed from the source signal needs to be transported and delivered to the intended users. We propose a novel implicit semantic communication (iSC) architecture for representing, communicating, and interpreting the implicit semantic meaning. In particular, we first propose a graph-inspired structure to represent implicit meaning of message based on three key components: entity, relation, and reasoning mechanism. We then propose a generative adversarial imitation learning-based reasoning mechanism learning (GAML) solution for the destination user to learn and imitate the reasoning process of the source user. We prove that, by applying GAML, the destination user can accurately imitate the reasoning process of the users to generate reasoning paths that follow the same probability distribution as the expert paths. Numerical results suggest that our proposed architecture can achieve accurate implicit meaning interpretation at the destination user. 
\end{abstract}
\vspace{-0.2in}

\section{Introduction}
Recent developments in communication networking systems have witnessed a growing interest in human-oriented services and applications such as VR/AR/XR and Tactile Internet, most of which are data-hungry and resource-consuming. The traditional content-agnostic data-driven communication architecture is now viewed as a major obstacle for delivering quality-of-experience (QoE)-demanding services to end-users. This motivates a novel paradigm, referred to as  {\em semantic communication},  which allows the meaning of messages to be identified and utilized during communication. Compared to the existing data-oriented communication networks, semantic communication allows all the communication participants including both information source and destination users to exploit commonly-shared human knowledge and experience as well as syntax, semantics, and inference rules to assist the transportation and accurate delivery of the intended meaning. Recent observation suggests that semantic communication has the potential to significantly improve efficiency and reliability of communication, enhance users' QoE, and achieve smoother cross-protocol/domain communication\cite{XY2021SemanticCommMagazine,XY20206GSelfLearn}.


Most existing works in semantic communication focused on transporting the explicit semantic information, e.g., the labels of objects that can be directly identified from the source signals (e.g, image, voice, and text signals) using mature AI algorithms. For example, in \cite{guler2018semantic}, the authors interpreted the semantics as each individual word identified from the message. Similarly, the authors in \cite{xie2020lite} defined the semantic of the source data as the meaning of a text. Explicit semantic information can be directly recognized from various forms of source signals (e.g., image, voice, text, etc.) by adopting mature machine learning approaches, such as regression and  deep-learning-based classification and recognition solutions. 

It is known that the information that can be communicated between users is much more than just explicit information. For example, an image showing ``a kid is riding a bicycle" consists of explicit objects ``a kid" and ``a bicycle". The relationship (``ride") between these two objects however cannot be directly recognized from the image. In another example, a child sends a voice message asking her father ``what is a Tweety". The key semantic element of this message ``Tweety" can have various interpretations including a smartphone App of a social media website, a canary bird, and a character in a cartoon TV show. To interpret the exact meaning of the message, the receiver (the father),  in this case must be able to infer the implicit information from the context and background of the child.
	
From the above examples, we can observe that, in addition to explicit information, the content of communication often consists of rich implicit information that is very difficult to represent, recognize, or recover\cite{XY2021SemanticCommMagazine}. Most of the existing works in semantic communication ignore implicit semantic due to the following reasons. First of all, there is lacking a simple and comprehensive way to represent implicit semantic. Different from explicit semantic that can be directly recognized from the source signal based on the labels of objects, the implicit meaning often involves many unobservable relations and hidden concepts that cannot be directly identified from the source signal. Secondly, The implicit meaning is also difficult to infer and can be closely related to user-related information. In other words, when observing the same source signal, different users can have different interpretations due to the difference in users' preference, personality, and background, e.g., when asking about the word ``Tweety", different kids may refer to different concepts (bird or cartoon character). Finally, accurately recovering and evaluating the implicit meaning at the destination user is also known to be a challenging task. Most existing works assume that the destination user can have a well-formulated analytic expression e.g., a reward or utility function, that can be directly optimized to maximize its understanding of semantic meaning of the source signal which is unrealistic in most practical scenarios.

In this paper, we propose a novel implicit semantic communication architecture, called iSC, for representing, modeling, and optimizing the delivery of implicit meaning of message. In particular, we first adopt a graph-based semantic representation, called semantic graph, which includes three key components: entity (objects and abstract concepts), relations (connections between entities), and reasoning mechanism (user's  reasoning and inference preference). {Our proposed representation is comprehensive enough to include both common knowledge shared among users and the personal reasoning preference as well as some private knowledge terms of each individual user.} We then introduce a novel generative imitation learning-based reasoning mechanism learning solution, called GAML, for supporting automatic encoding, transportation, and decoding/interpretation of implicit semantic. In this solution, the encoder (at the source user) will assist the decoder (at the destination user) to train a reasoning mechanism to automatically map explicit objects identified from the source signal to a set of possible hidden concepts and objects that are relevant to the semantic meaning. Motivated by the recent discovery that human users tend to reason hidden concepts and ideas by following their directly linked knowledge entities and relations, we approximate the reasoning process of each user as a Markov decision process (MDP) in which all the hidden entities and relations involved in implicit semantic are discovered sequentially from the explicit objects. To address the issue that, in most MDP-based problems, deriving the optimal policy often requires the reward functions to be analytically expressed, we employ a generative imitation learning-based approach for the decoder to learning and imitate the reasoning process of the source users without knowing nor modeling any specific reward function. We prove that by applying GAML, the decoder will learn a reasoning mechanism to generate reasoning paths from explicit objects that have the minimum statistic difference, i.e., semantic distance, to the expert paths observed by the source user. %

We summarize  main contributions of this paper as follows:

\noindent{\bf New graphical representation of semantic meaning:}
 { a novel solution for representing implicit semantic is proposed. Our proposed representation consists of three key components: entities, relations, and reasoning mechanism. We have shown that the proposed three-component structure can provide a comprehensive way for representing implicit meaning taking into account both commonly shared knowledge as well as user-related private information such as personal preference in reasoning and inference.  }

\noindent {\bf Novel implicit semantic communication architecture:} We propose a novel architecture, iSC, supporting automatic encoding and decoding of implicit semantic at the source and destination users, respectively. Our proposed architecture does not require the destination users to observe any expert reasoning results at the source user nor transmitting of any implicit semantic information from the source to the destination users.

\noindent {\bf New reasoning mechanism learning solution:} We introduce a novel generative adversarial imitation learning-based solution, GAML, for the decoder to learn the reasoning mechanism of the source user. We prove that by adopting GAML, the decoder can automatically generate implicit reasoning paths that follow the same probability distribution as the expert reasoning paths observed by the source user.

 \noindent{\bf Extensive simulations:} We conduct extensive experiments based on real-world dataset. Our results suggest that the proposed solution can accurately recover implicit reasoning results. 

115

	\vspace{-0.2in}
	\section{Primer}
	\label{Section_Primer}
	\subsection{Representation of Semantic Knowledge}
	One of the key issues in semantic communication is to develop a general and comprehensive way to represent the meaning of message. In this paper, we propose a graph-based representation of semantic meaning that can cover both explicit and implicit meaning. Our proposed semantic representation consists of three key components:
	
\noindent{\bf Entities:}
	correspond to real-world objects and concepts, such as ``kid" ,``bicycle", ``Tweety", ``social media website", etc.
	
\noindent{\bf Relations:}
	represent the relationship between entities, e.g, a kid ``rides" a bicycle, a Tweety ``is" a canary bird, and a Tweety ``is" a cartoon character ``in" a TV show.
	
\noindent{\bf Reasoning Mechanism:}
	In addition to the entities and relations, the meaning of a message may also include a reasoning mechanism capturing possible relations and hidden entities that cannot be directly identified from the source signal. For example, in the previous example, the entity ``Tweety" may also link to several other hidden entities such as ``smart phone App", ``canary bird" and ``cartoon character" with corresponding hidden relations. The receiver (the father) therefore needs to infer possible entities and the corresponding relations that link to entity ``Tweety". For example, if the kid recently reads some books about animal, the father would be able to infer the most appropriate meaning of the kid's question by finding a path ``Tweety is a canary bird that is highly likely to be shown in a book recently read by the kid". It can be observed that the reasoning mechanism plays an essential role in inferring the appropriate meaning of the message. Generally speaking, the semantic of a source signal can involve at least one key entity and the reasoning paths associated with these key entities may not be unique and can be closely related to the user's background, environment, and context of message.
	
	To better interpret the meaning of these entities, the key issue is to learn the optimal reasoning mechanism that can output relevant reasoning paths based on the identified entities. 
	\vspace{-0.2in}
	\subsection{Semantic Knowledge Base}
 We refer to the collection of all the  knowledge entities and relations that are accessible for each user as the semantic knowledge base. Generally speaking, the knowledge base of each user consists of both common knowledge shared among users as well as some private knowledge that is only available at each individual user. 
	
Our proposed semantic knowledge base is closely related to the concept of knowledge graph (KG) with the following key differences. First, KG is built based on real-world facts and words with meanings carefully defined by the linguists while ignoring some private users' personal preference, experience, and incorrect understanding of concepts. For example, WordNet \cite{miller1995wordnet}, one of the most popular KGs, is built 
based on the meanings defined in a dictionary. Similarly, the relations in the KG is also defined based on the fact-based relationship, e.g., the synonymous and antonymous defined in thesaurus.
	
In this paper, we focus on the semantic communication from a source user to a destination user, labeled as $E$ and $D$, respectively. We use source user (or destination user) and encoder (or decoder), interchangeably.  We assume the source and destination have already established their knowledge bases, denoted as ${\cal{K}}^{E} = \langle {\cal{E}}^{E}, {\cal{R}}^{E} \rangle$ and ${\cal{K}}^{D} = \langle{\cal{E}}^{D}, {\cal{R}}^{D}\rangle $. The knowledge base of the user can be located and cached in the user's device memory or stored at the closest edge server.
	
	\vspace{-0.2in}
	\subsection{Reasoning Mechanism Modeling and Learning}
Understanding the implicit meaning that cannot be directly identified from the source signal is known to be a notoriously challenging task. In this paper, we propose a novel solution, GAML, based on imitation learning to infer the implicit relational path from the key entity towards the most probable hidden entities. Our proposed solutions are inspired by existing KG reasoning solutions focusing on learning and imitating the reasoning trajectories obtained from previous observations, known as the expert reasoning paths. Unfortunately, the existing KG reasoning solutions cannot be directly applied to the implicit semantic communication due to the following reasons:

\noindent{(1)} {Knowledge triplet commonly investigated in KG reasoning cannot always be  identified from the source signal.} In particular,
Most existing multi-hop reasoning solutions focus on predicting the missing relation or entity of a carefully defined knowledge triplet. 
However, in semantic communication, it is generally impossible to identify the exact subject, predicate, or object from any message.
		
\noindent{(2)} {In many existing KG reasoning solutions, the rewards of successful reasoning must be analytically expressed, which is unrealistic in many practical scenarios. } 

\noindent{(3)} {There is still lacking a commonly adopted metric for measuring the semantic distance between implicit semantic meanings of different signals in communication systems.}

	\begin{figure}
		\centering
		\includegraphics[width=7cm]{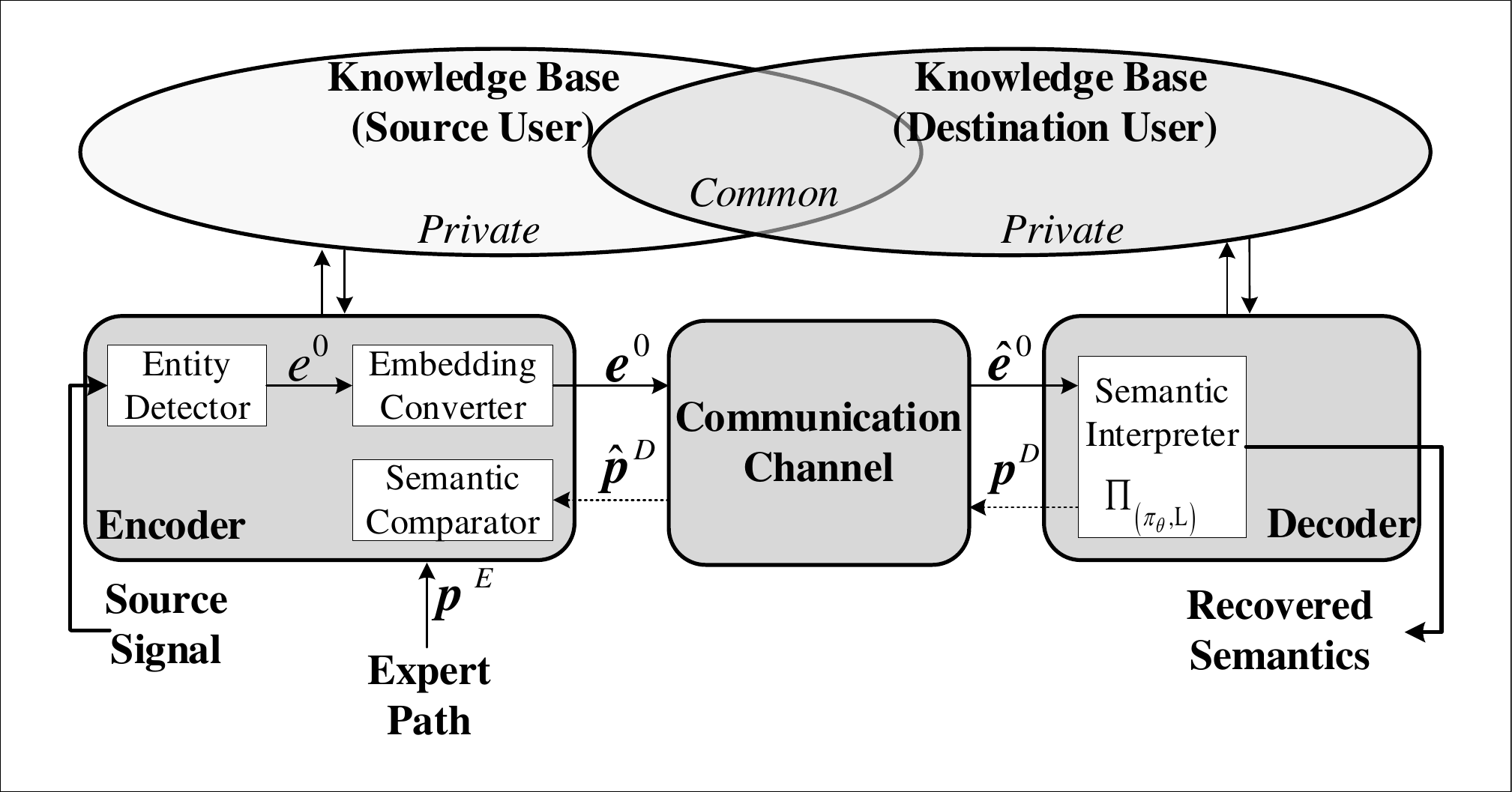}
		\caption{The proposed iSC architecture.}
		\label{Fig_systemmodel}
		\vspace{-0.2in}
	\end{figure}

\vspace{-0.15in}
\section{System Model and Problem Formulation}
\label{Section_SystemModel}
\subsection{System Model}
We focus on semantic communication between a single pair of source and destination users consisting of the following key functional components as shown in Fig. \ref{Fig_systemmodel}.

\subsubsection{Semantic Encoder} consists of the following key  sub-components:

\noindent{\bf (Explicit) Entity Detector:} The encoder should be able to first identify one or more key entities, denoted as $e^{0}$, from various forms of the source signals. This can be achieved by pre-installing well-trained AI models such as YOLO \cite{redmon2016look} and WaveNet \cite{audio2019} to identify known labels of objects. 

\noindent{\bf Semantic Comparator:} One of the key differences between semantic communication and traditional communication is that the delivered result of a message is generally not binary 
but can be characterized by the {\it semantic distance}, a continuous value characterizing how far the meaning interpreted by the receiver diverges from the original meaning of the source user. Let  $\Gamma\left(\eta^E, \eta^D\right)$ be the semantic distance between the original meaning $\eta^{E}$ of source user and the meaning $\eta^D$ recovered by the destination user. 
In iSC, the semantic meaning should consist of a reasoning path including one or more sequences of entities and relations originated from the recognized key entities, i.e., $\eta^E = \langle{e^{0}, r^1, e^1, r^2, e^2, \ldots}\rangle$. 

\subsubsection{Semantic Decoder} consists of the following key functional sub-component:

\noindent{\bf Semantic Interpreter:} The decoder can recover a reasoning path $\eta^D$ representing its interpretation of the implicit meaning associated with the key entities. The destination user can only receive a noisy version  of key entities, denoted as ${\hat e}^0$, sent from the source user and will be able to learn a reasoning mechanism to output the possible reasoning paths $\eta^D=\langle{{\hat e}^0, {\hat r}^1, {\hat e}^1, {\hat r}^2, {\hat e}^2, \ldots}\rangle$.

\vspace{-0.15in}
\subsection{Problem Formulation}	
The main objective is to develop a novel solution for semantic decoder to automatically output a reasoning path $\eta^D$ that has the minimum semantic distance to the original meaning $\eta^E$ of the source signal, i.e., we try to solve the following problem
\begin{eqnarray}
\mbox{(P1)} \;\;\;\; \min_{\theta} \Gamma_{\theta}\left(\eta^{E}, \eta^D\right) \nonumber
\end{eqnarray}
where $\theta$ represents the latent parameters of the semantic interpreter at the decoder.

As mentioned earlier, due to the complexity of knowledge base and rich meanings that can be represented in the communication messages, it is generally difficult to find a simple and comprehensive approach to solve (P1). In this paper, we propose an imitation learning-based framework 
in which the source and destination users can learn from a set of reasoning paths, called expert paths, that can only observed by the source user.
We assume the reasoning mechanism of the user can be approximated by an MDP process and then our proposed learning framework allows the destination user to train a semantic interpreter to automatically construct a policy network to generate reasoning paths that match the distribution of the expert paths. 

\section{iSC Architecture}
\label{Section_Architecture}

\subsection{Semantic Encoder}
\noindent{\textbf{Semantic Distance:}}
Finding a proper metric to characterize the difference in meaning of two reasoning paths is of critical importance for semantic communication systems. In existing semantic communication solutions, the semantic distance is often measured by word similarity (e.g, defined by WordNet) between the labels of objects. 
Implicit semantic distance however cannot adopt the similarity of objects (words) to measure semantic distance due to the following reasons. First, entities and relations play different roles in interpreting the meaning of each reasoning path. Also, the output of the neural networks is the probability distribution of all the possible paths, i.e., the likelihood of paths composed of different combinations of relations and entities, and is, therefore, difficult to have a single value metric to characterize the meaning difference between two reasoning paths.

One way to solve this issue is to project the complex relations into a low-dimensional space called the embedding space and then adopt a certain distance metric, e.g., Euclidean distance, to measure the difference in meaning. Let $\be$ and $\br$ be the embeddings of entity $e$ and relation $r$, respectively.  
In this paper, we adopt a commonly used graph embedding solution, called TransE \cite{bordes2013translating}, due to its simplicity. In this embedding model, the addition of an entity embedding $\be^0_i$ and a relation embedding $\br^1_i$ will be close to the embedding of the connected entity embedding $\be^1_i$, i.e., $\be^0_i+\br^1_i\approx \be^1_i$. We can then convert a reasoning path $\eta  = \langle e^0, r^1, e^1, r^2, e^2,\ldots\rangle$ into the following path embedding $\bp = \sum_{i}{\br}^i$.

Let $\bp^E$ and $\bp^D$ be the expert path embedding and the generated path embedding by the decoder, respectively. Without loss of generality, we assume $\bp^E$ and $\bp^D$ have the same dimension size. 
We can then feed the path embedding into a neural network to obtain the extracted semantic feature of expert paths and generated paths as follows:
\begin{eqnarray}
\varpi_{\phi}\left({\bp}^E\right) &=& \sigma\left(\omega_2, f\left(\omega_1, {\bp}^E\right)\right) \label{eq_varpi1} \\
\varpi_{\phi}\left({\bp}^D\right) &=& \sigma\left(\omega_2, f\left(\omega_1, {\bp}^D\right)\right) \label{eq_varpi2}
\end{eqnarray}
where $f\left(\cdot\right)$, $\sigma\left(\cdot\right)$ are activation functions (e.g., ReLU, softmax, sigmod, etc.) at various layers of neural networks, i.e., in (\ref{eq_varpi1}) and (\ref{eq_varpi2}), we consider a two-layer neural network in which $\omega_1$  and $\omega_2$ are the weights of the first layer and second layer, respectively.
Latent parameter $\phi$ of the comparator $\Gamma$ consists of all the parameters of these layers as well as the activation functions adopted at different layers.
We then define the difference of extracted semantic features of two reasoning paths as their {\em semantic distance}, i.e., the semantic distance between meanings represented by two reasoning paths $\bp^E$ and $\bp^D$ is given by
\begin{eqnarray}
\Gamma \left(\bp^E, \bp^D\right) = \varpi_{\phi}\left({\bp}^E\right)- \varpi_{\phi} \left({\bp}^D\right)
\label{eq_SemanticDistance}
\end{eqnarray}



\noindent{\textbf{Semantic Comparator:}} During the training  stage, the encoder can collect a set of expert paths $\mathcal{T}^E$. These expert paths $\mathcal{T}^E$ can be obtained by previous observations and/or sampling from the source user's knowledge base. We follow a commonly adopted setting and assume the expert paths follow a certain distribution, denoted as $\Delta^E$. During the training, the encoder will identify some key entities  $e^0$ 
from the expert paths $\mathcal{T}^E$ to be sent to the semantic interpreter at the decoder for generating some random paths $\mathcal{T}^D$ based on the destination user's knowledge base $\mathcal{K}^D$. These generated reasoning paths $\mathcal{T}^D$ will be feedback to the semantic comparator at the encoder. The semantic comparator at the source user will calculate the semantic distance between the generated paths $\mathcal{T}^D$ and expert paths $\mathcal{T}^E$ to be sent to the decoder. In this way, the semantic interpreter and semantic comparator will be interactively trained by minimizing the semantic distance between the guaranteed reasoning paths $\mathcal{T}^D$ and the expert paths $\mathcal{T}^E$. 
More formally, 
suppose $\bp^E$ and $\bp^D$ are two embeddings corresponding to expert paths and paths generated by the decoder. The semantic comparator will be trained to  better differentiate the semantic meaning of expert paths and paths generated by the decoder, i.e., if we adopt a commonly used loss function, cross entropy, as our loss function, we can write the optimization problem for semantic comparator as follows:
\begin{eqnarray}
\underset{\varpi_{\phi}}{\max}\left( \mE [{\log \varpi_{\phi}\left(\bp^E\right)}]-\mE [1- \log \varpi_{\phi}\left(\bp^D\right)]\right)
\label{eq_comparator}
\end{eqnarray}

\vspace{-0.3 in}

\subsection{Semantic Decoder}

\noindent{\bf Semantic Interpreter:} Due to the complexity and diversity of implicit reasoning results, it is unrealistic to assume the reasoning path generated from any key entity is unique and deterministic. In fact, the meaning represented by different reasoning paths can be very similar and therefore it is generally difficult to determine a single path that best interprets the meaning of the source signal. A more realistic and practical solution is to learn a reasoning mechanism based on reinforcement learning to map the received key entities into the probability distribution of multiple possible reasoning paths.  In this paper, we follow a commonly adopted setting and assume the relational reasoning process from each entity in a given knowledge base can be formulated as an MDP process in which the main objective is to learn a policy that can specify the probability distribution of choosing a set of connected relations to extend the reasoning path when arriving at each entity. In this way, the semantic interpreter can output all the possible reasoning paths by extending the possible relations when arriving at each explicit or hidden entity originated from one or more key entities. More formally, we define the relational reasoning process as a MDP $\langle {\cal A}, {\cal S}, R, {\Gamma} \rangle$ consisting of the following components:

    \noindent \textbf{(a) State:} In an MDP-based reasoning, the reasoning path is generated by choosing one relation and entity pair at a time. As mentioned earlier, the encoder will convert the entities and relations into an embedding space in which the distance between entities extended from different paths of relations reflects the closeness of the represented meanings. In this case, the state includes the embedding of the current entity as well as the original key entity received from the encoder, i.e., the state at the $t-$th iteration of path reasoning is given by $\bs_t = \left(\be_t, \be_0\right)$, where $\bs_t \in {\cal S}$ and ${\cal S}$ is the state space.   

    \noindent \textbf{(b) Action:} Action space $\cal A$ is defined as the set of all the possible relations specified in the user's knowledge base. More formally, given the current state $s_t$, the action of the user is to choose the possible relations to extend the paths. 

    \noindent \textbf{(c) Reward:} The main objective is to minimize the semantic distance between the expert paths observed by the encoder and the reasoning paths generated by the semantic interpreter at the decoder. 

 \noindent \textbf{(d) Policy:} We define the policy network as a neural network parameterized by $\theta$. Our policy network maps the current state $\bs_t$ into a probability distribution over all the possible relations to extend the path. For example, when adopting a three layer neural network, the output of policy network is given by
 \begin{eqnarray}
\pi_{\theta} \left( \bs_t \right) = {\sigma}\left(\omega_3', g\left(\omega_2',f(\omega_1',\bs_t)\right)\right)
 \end{eqnarray}
	where $\sigma\left(\cdot\right)$, $g\left(\cdot\right)$, $f\left(\cdot\right)$ are activation functions and $\omega_1'$ , $\omega_2'$, $\omega'_3$ are weights of the first, second, and third layers, respectively.

Note that the policy network only specifies the probability distribution of relations extended from a given entity. The decoder however needs to recover the full reasoning path consisting of a sequence of entities and relations.
We therefore define a reasoning mechanism $\Pi_{\left(\pi_{\theta},L\right)}$ which maps a policy network into a probability distribution of all the possible paths under a certain constraint $L$. The constraint $L$ can be closely related to the depth of meaning that can be expressed by the user. For example, a shorter reasoning path generally represents a relatively more straightforward meaning. As the length of the path continues to grow, the chance of disclosing some deep meaning increases. However, it will also increase the searching space and chance to misrepresent the real meaning of the source user.
More formally, the main objective of the semantic interpreter can be written as follows:
\begin{eqnarray}
\lefteqn{\underset{\pi_\theta}{\min} \left( \mE_{\bp^E \sim \Delta^E} [{\log \varpi^*_{\phi}\left(\bp^E\right)}] \right.}\nonumber\\
&&\;\;\;\;\; \left.-\mE_{\bp^D \sim \Pi_{\left(\pi_{\theta},L\right)}} [1- \log \varpi^*_{\phi}\left(\bp^D\right)]\right)\label{eq_interpreter}
\end{eqnarray}
where $\bp ^D$ is the  embedding of path $\eta^D$ and $\varpi^*_{\phi}$ is the solution of (\ref{eq_comparator}).

\vspace{-0.2in}
\subsection{Theoretical Analysis}

\begin{algorithm}
\caption{GRML Algorithm}
\label{Algorithm_JointTraining}
{\bf Input}: Key entity $e^0$, expert semantic paths set $\mathcal{T}^E$, initial policy network $\pi(\bs|\theta_0)$ and comparator network $\varpi(\bp|\phi)$ with initial parameters $\theta_0, \phi_0$, and max length of hops $L$\\
{\bf Output} learned policy $\pi^*$\\
{\textbf{For}} number of training iterations {\textbf{do}}
\begin{itemize}
    \item  Generate semantic paths set $\mathcal{T}_i^D$ from policy $\pi_{\theta_i}$

    \item  Update comparator parameters $\phi_i$ to $\phi_{i+1}$ with the following gradient:
    \begin{equation}
    \centering
    \begin{aligned}
     &\mathbb{E}_{\bp_j^D \sim \Pi_i^D}[\nabla _{\phi}\log \varpi_{\phi}(\bp_j^D)] +\\&  \mathbb{E}_{\bp_j^E \sim \Delta_i^E}[\nabla _{\phi}(1-\log \varpi_{\phi}(\bp_j^E))]
    \end{aligned}
    \end{equation}
    \item   Update policy parameters $\theta_i$ to $\theta_{i+1}$ by minimizing the cost function with Monte Carlo Policy Gradient:
    \begin{eqnarray}
        &\mathbb{E}_{\bp_{j}^D\sim \Pi^D}[\nabla_{\theta}\log(\pi_{\theta}(\ba|\bs)Q(\bs, \ba))]-\alpha\nabla_{\theta}H(\pi_\theta),
    \end{eqnarray}
where $Q(\bs, \ba) = \log (\varpi_{\phi_{i+1}}(\bp_j^D))$.
    \end{itemize}
{\textbf{End for}}
\end{algorithm}

We summarize the main training procedures in Algorithm  \ref{Algorithm_JointTraining}. We can prove the following result.
\begin{theorem}
\label{MainConvergTheorem}
Suppose at each iteration of GAML Algorithm, the semantic comparator can achieve its optimal solution $\varpi^*_\phi$ of (\ref{eq_comparator}) under given $\pi_\theta$ and $\Pi_{\left(\pi_{\theta},L\right)}$ is updated according to $\pi_\theta$. We can then prove that when $\pi_\theta$ converges to the optimal solution $\pi^*_\theta$ of (\ref{eq_interpreter}), the distribution $\Pi_{\left(\pi_{\theta},L\right)}$ of generated semantic paths at the decoder approaches to the distribution $\Delta^E$ of expert paths.
\end{theorem}
\begin{IEEEproof}
It can be observed that our proposed architecture is very similar to a modified generative adversarial networks (GANs) fitted into a semantic communication scenario. We omit the details due to the limit of space. 
\end{IEEEproof}

\vspace{-0.1 in}
\section{Experiments}
\label{Section_NumericalResult}

\subsection{Dataset and Simulation Setup}
We use NELL-995, a large real-world knowledge dataset\cite{xiong2017deeppath}, to evaluate the performance of our proposed iSC architecture and the GAML algorithm.
NELL-995 consists of $754,920$ unique entities and $200$ types of relations. We  sample a set of expert paths from NELL-995 dataset using a two-side breadth first search algorithm to simulate the source user who tends to express his/her meaning based on two-hop reasoning paths. We then adopt TransE\cite{bordes2013translating} to convert these expert paths into a 100-dimensional continuous embedding space. We set the semantic interpreter as a fully-connected network consisting of two hidden layers, each followed by a Rectified Linear Unit (ReLU) and one output layer. The output of the interpreter is normalized using a softmax function. For the semantic comparator, we adopt a two-layer fully-connected network with one hidden layer and one output layer. The output layer is normalized by a sigmoid function while others are activated by ReLU.
Our simulations are performed based on Tensorflow open source platform on a workstation with an Intel(R) Core(TM) i9-9900K CPU@3.60GHz, 128.0 GB RAM@2133 MHz, 2 TB HD, and two NVIDIA Corporation GP102 [TITAN X] GPUs.

\vspace{-0.2in}
\subsection{Numerical Results}

\begin{figure}
  \begin{minipage}[t]{0.49\linewidth}
   \centering
   \includegraphics[width=4.3cm]{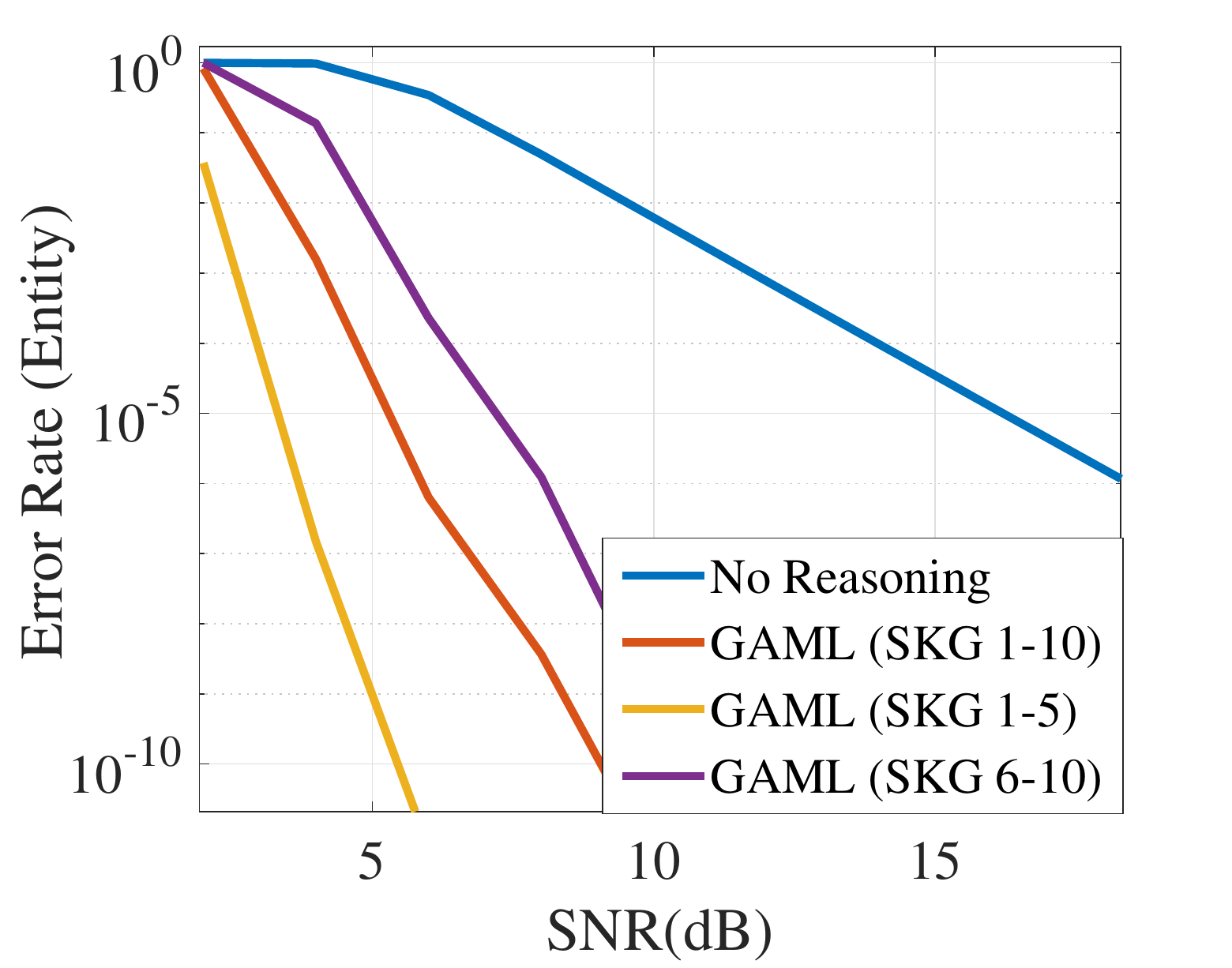}
   \vspace{-0.15in}
	 \caption{\small{Packet (entity) error rate with and without semantic-reasoning under different SNRs.}}
   \label{Fig_SNR}
  \end{minipage}
  \begin{minipage}[t]{0.49\linewidth}
	\centering
	\includegraphics[width=4.3cm]{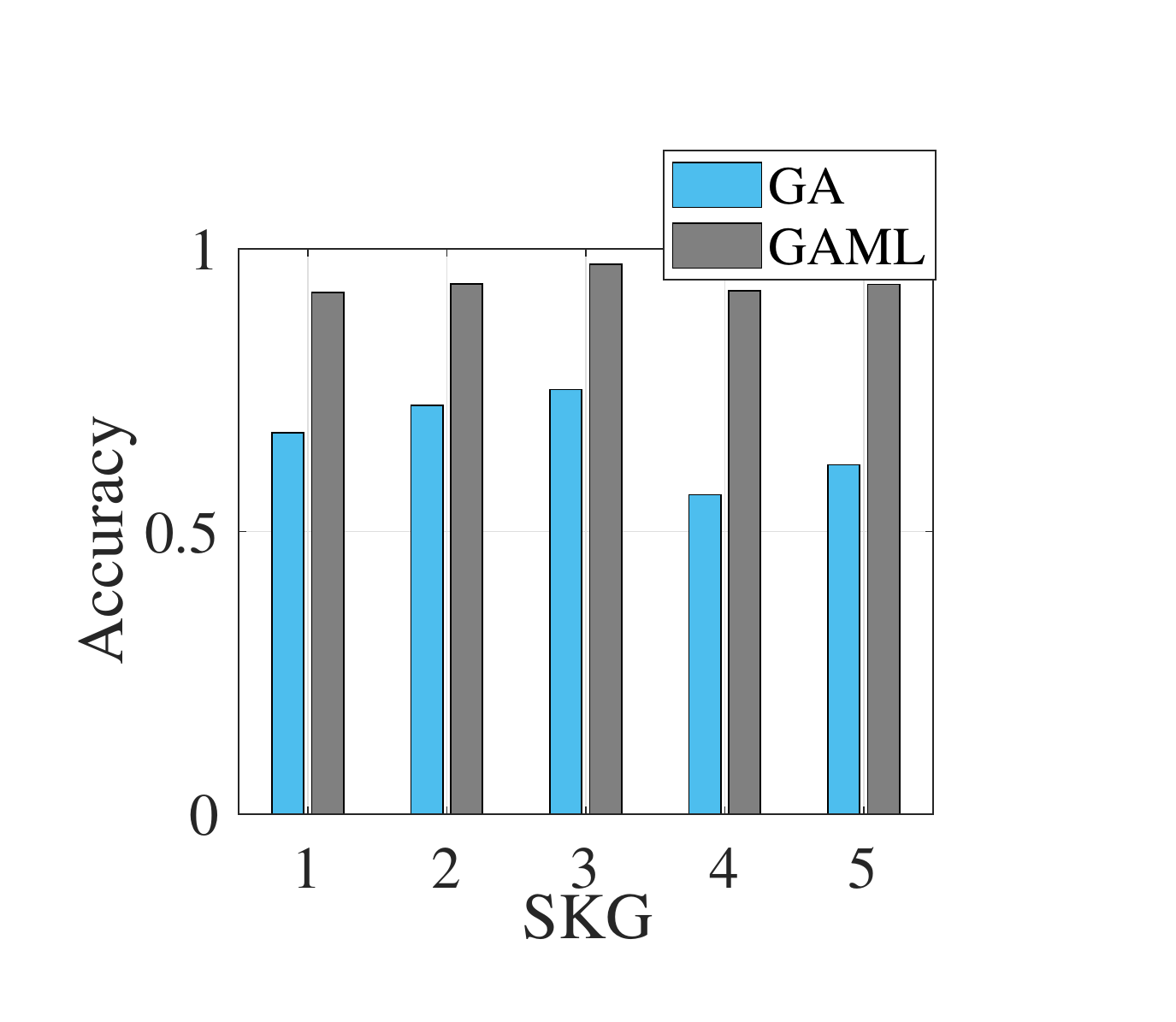}
	\vspace{-0.15in}
	\caption{\small{Accuracy of semantic path prediction achieved by GAML compared with GA under different SKGs.}}
	\label{Fig_Accuracy}
  \end{minipage}%
  \vspace{-0.2in}
 \end{figure}
\begin{figure}
    \begin{minipage}[t]{0.5\linewidth}
	\centering
	\includegraphics[width=4.4cm]{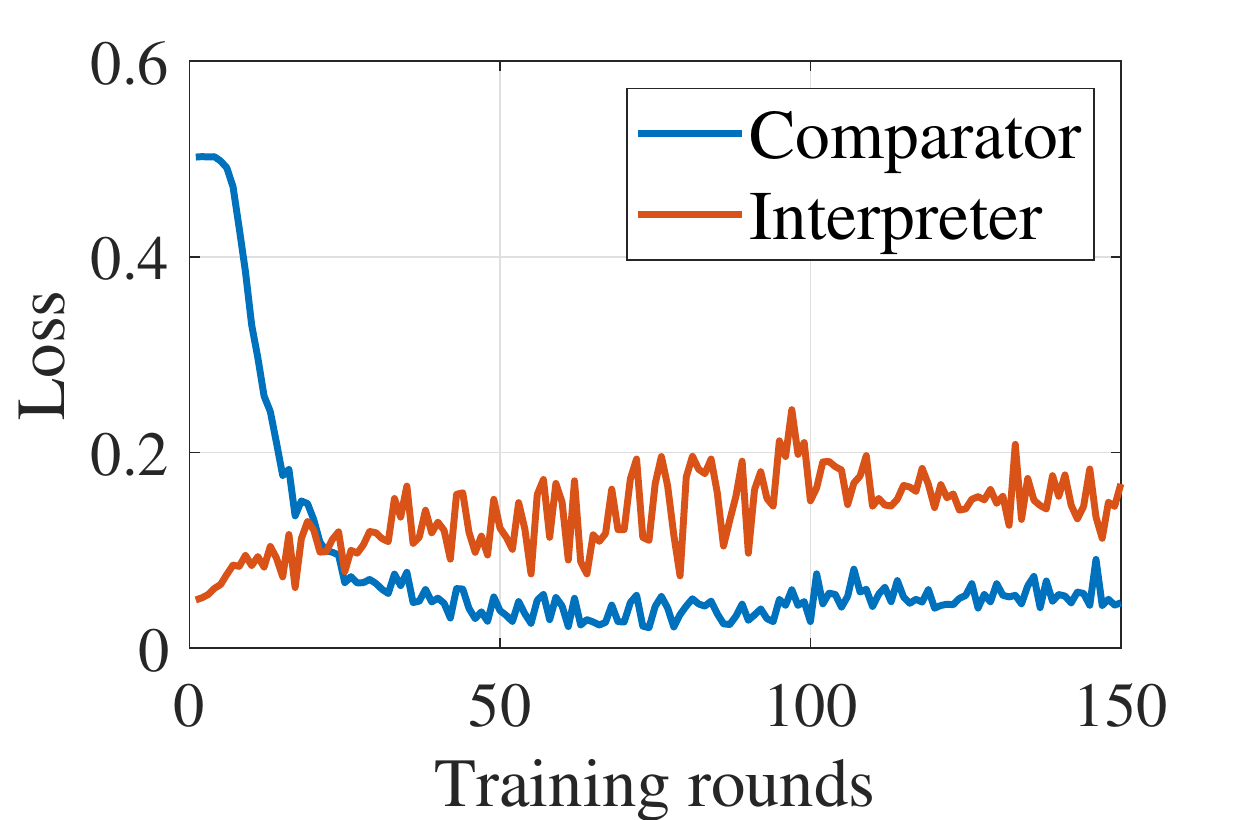}
	\vspace{-0.15in}
	\caption{\small{Loss of semantic \\comparator and interpreter\\ under different training rounds.}}
	\label{Fig_convergence}
  \end{minipage}%
  \begin{minipage}[t]{0.45\linewidth}
	\centering
	\includegraphics[width=4.4cm]{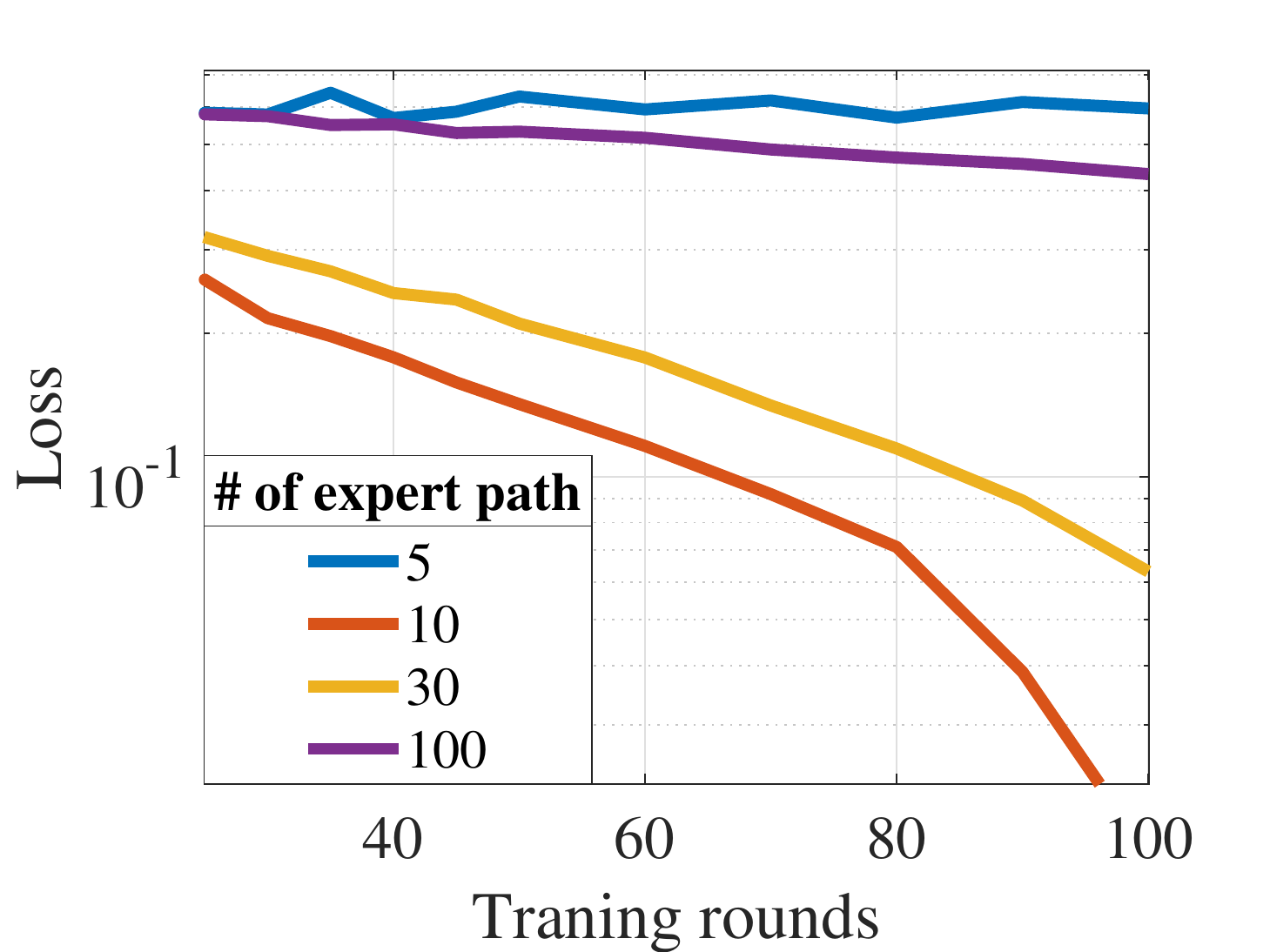}
	\vspace{-0.15in}
	\caption{\small{Loss when being trained with different numbers of expert paths.}}
	\label{Fig_NumExpertPath}
  \end{minipage}%
  \vspace{-0.2in}
\end{figure}

We first evaluate the potential communication performance that can be improve by using our proposed semantic reasoning mechanism in Fig. \ref{Fig_SNR} where we present the data packet loss of an additive Gaussian noise channel when the packets corrupted by the physical channel can be recovered by using our proposed reasoning-based solutions. Since in our considered dataset, the number of relation types is much less than the number of entities, it is therefore unnecessary for the source user to always send all the embedding feature sets of each relation in each transmission. The source user can send the full embedding data of each type of relation once and then use a simple label to denote the type of relations during the rest of communication. We therefore mainly focus on the entity recovery performance when a source user sends all the entities, each represented by a packet with size of 3800 bits, to the destination user. We can observe that our proposed reasoning-based solutions can achieve a significant error correction performance, compared to the traditional communication solution without semantic reasoning.
As mentioned earlier, the accuracy of semantic reasoning is also closely related to the structural feature of the knowledge graph. For example, the density of connections among entities may reflect the similarity of meaning of these entities, i.e., the higher the density of connections, the closer the meaning among these entities. To investigate impact of meaning similarity of entities on the error correction performance, we consider 10 sub-knowledge graphs (SKG), each consists of an exclusive set of $75,492$ entities linked with different number of relations and reasoning paths. We then rank these 10 SKGs according to their connection densities from the highest to the lowest to simulate the communication involving entities with different levels of semantic meaning diversity, i.e., semantic diversities of SKGs 1 and 10 are the lowest and the highest, respectively. We present the packet error rate achieved by our proposed reasoning-based error recovery solution with different combinations of SKGs in Fig. \ref{Fig_SNR}. We can observe that our proposed solution achieves a higher error correction performance for entities with lower semantic diversity.

We compare our proposed GAML with genetic algorithm (GA)-based reasoning solution, a popular heuristic knowledge inference solution based on population evolution and competition\cite{zhang2017maximizing} in Fig. \ref{Fig_Accuracy}. We compare the semantic path reasoning performance of our proposed GAML and GA under different SKGs. We can observe that both our proposed GAML and GA solution are influenced by the semantic diversity of entities and GAML offers at least 20\% of performance improvement over GA in all 5 considered SKGs.

To verify the convergence of our proposed mechanism learning solution, in Fig. \ref{Fig_convergence}, we present the loss values of semantic comparator and interpreter under different training rounds defined in (\ref{eq_comparator}) and (\ref{eq_interpreter}), respectively. We can observe that the loss functions of both comparator and interpreter can approach relatively stationary values with only 50 rounds of training. This means that the communication overhead for training a relatively satisfactory model at the semantic interpreter is moderate.

In Fig. \ref{Fig_NumExpertPath}, we evaluate  the convergence of our proposed reasoning mechanism learning solution in Algorithm \ref{Algorithm_JointTraining} when the model has been trained with different numbers of expert paths. We can observe that when the number of expert paths is limited, the proposed algorithm cannot converge. The convergence performance improves when the number of observable expert paths increases. This is because, the overall semantic diversity of training paths decreases with the number of training  paths, e.g., the chance of having semantic similar paths increases with the total number of training paths, which   results in faster convergence rate especially when the number of training rounds increases. 

\vspace{-0.15 in}
\section{Conclusion}
\label{Section_Conclusion}
This paper proposed a comprehensive framework for representing, modeling, and interpreting implicit semantic meaning among users. We first introduced  a novel graph-inspired structure to represent the implicit meaning of message  
and then developed a novel semantic communication architecture, iSC, in which a reasoning mechanism can be trained at the destination user with the help of the source user. 
A generative imitation learning-based framework was then introduced for the destination user to imitate the reasoning process observed by the source users. We  proved that, by applying our proposed learning framework, the decoder will learn a reasoning mechanism to generate reasoning paths that follow the same probability distribution as the expert paths. Numerical results suggest that our proposed architecture achieves accurate implicit meaning interpretation at the destination user with limited communication overhead.

\section*{Acknowledgment}
This work was supported in part by the National Natural Science Foundation of China under Grant No. 61836008 and 62071193, the Pengcheng National Laboratory project under Grant No. PCL2021A12, and the Key R\&D Program of Hubei Province of China under Grant No. 2021EHB015 and 2020BAA002.


\bibliographystyle{IEEEtran}
\bibliography{IEEEabrv,mybib}

\begin{thebibliography}{10}
\providecommand{\url}[1]{#1}
\csname url@samestyle\endcsname
\providecommand{\newblock}{\relax}
\providecommand{\bibinfo}[2]{#2}
\providecommand{\BIBentrySTDinterwordspacing}{\spaceskip=0pt\relax}
\providecommand{\BIBentryALTinterwordstretchfactor}{4}
\providecommand{\BIBentryALTinterwordspacing}{\spaceskip=\fontdimen2\font plus
\BIBentryALTinterwordstretchfactor\fontdimen3\font minus
  \fontdimen4\font\relax}
\providecommand{\BIBforeignlanguage}[2]{{%
\expandafter\ifx\csname l@#1\endcsname\relax
\typeout{** WARNING: IEEEtran.bst: No hyphenation pattern has been}%
\typeout{** loaded for the language `#1'. Using the pattern for}%
\typeout{** the default language instead.}%
\else
\language=\csname l@#1\endcsname
\fi
#2}}
\providecommand{\BIBdecl}{\relax}
\BIBdecl

\bibitem{XY2021SemanticCommMagazine}
G.~Shi, Y.~Xiao, Y.~Li, and X.~Xie, ``From semantic communication to
  semantic-aware networking: Model, architecture, and open problems,''
  \emph{IEEE Commun. Magazine}, vol.~59, no.~8, pp. 44--50, Aug. 2021.

\bibitem{XY20206GSelfLearn}
Y.~Xiao, G.~Shi, Y.~Li, W.~Saad, and H.~V. Poor, ``Towards self-learning edge
  intelligence in 6{G},'' \emph{IEEE Commun. Magazine}, Dec. 2020.

\bibitem{guler2018semantic}
B.~G{\"u}ler, A.~Yener, and A.~Swami, ``The semantic communication game,''
  \emph{IEEE TCCN}, vol.~4, no.~4, pp. 787--802, Sep. 2018.

\bibitem{xie2020lite}
H.~Xie and Z.~Qin, ``A lite distributed semantic communication system for
  internet of things,'' \emph{IEEE J-SAC}, vol.~39, no.~1, Jan. 2021.

\bibitem{miller1995wordnet}
G.~A. Miller, ``Wordnet: a lexical database for english,'' \emph{Communications
  of the ACM}, vol.~38, no.~11, pp. 39--41, 1995.

\bibitem{redmon2016look}
\BIBentryALTinterwordspacing
J.~Redmon, S.~Divvala, R.~Girshick, and A.~Farhadi, ``You only look once:
  Unified, real-time object detection,'' \emph{arXiv}, May. 2016. [Online].
  Available: \url{https://arxiv.org/abs/1506.02640}
\BIBentrySTDinterwordspacing

\bibitem{audio2019}
H.~Purwins, B.~Li \emph{et~al.}, ``Deep learning for audio signal processing,''
  \emph{IEEE J-STSP}, vol.~13, no.~2, p. 206–219, May. 2019.

\bibitem{bordes2013translating}
A.~Bordes, N.~Usunier \emph{et~al.}, ``Translating embeddings for modeling
  multi-relational data,'' in \emph{NIPS}, Nevada. Dec. 2013.

\bibitem{xiong2017deeppath}
W.~Xiong \emph{et~al.}, ``Deeppath: A reinforcement learning method for
  knowledge graph reasoning,'' \emph{arXiv 1707.06690}, Jul. 2017.

\bibitem{zhang2017maximizing}
K.~Zhang, H.~Du, and M.~W. Feldman, ``Maximizing influence in a social network:
  Improved results using a genetic algorithm,'' \emph{Physica A: Statistical
  Mechanics and its Applications}, vol. 478, pp. 20--30, 2017.

\end{thebibliography}
\end{document}